\documentclass[5p,sort&compress,longtitle]{elsarticle}  
\journal{Physics Letters B}
\usepackage{pifont,natbib,geometry,fleqn,graphicx}   
\usepackage{dcolumn}        
\usepackage{bm}               
\usepackage{amssymb}      
\usepackage{textcomp}
\usepackage{verbatim}
\usepackage{amsmath}
\usepackage{appendix}
\usepackage{enumitem}
\usepackage{epstopdf}
\usepackage{lineno}
\usepackage[utf8]{inputenc}  

\setdescription{leftmargin=4\parindent,labelindent=2\parindent}

\def\1/2{\frac{1}{2}}
\def\3/2{\frac{3}{2}}

\def\para{\parallel}

\begin{document}
\begin{frontmatter}

\title{Photon beam asymmetry $\Sigma$ in the reaction
$\vec{\gamma} p \to p \omega$ for $E_\gamma$ = 1.152 to 1.876 GeV}

\newcommand*{\ANL}{Argonne National Laboratory, Argonne, Illinois 60439}
\newcommand*{\ANLindex}{1}
\newcommand*{\ASU}{Arizona State University, Tempe, Arizona 85287-1504}
\newcommand*{\ASUindex}{2}
\newcommand*{\CSUDH}{California State University, Dominguez Hills, Carson, CA 90747}
\newcommand*{\CSUDHindex}{3}
\newcommand*{\CANISIUS}{Canisius College, Buffalo, NY}
\newcommand*{\CANISIUSindex}{4}
\newcommand*{\CMU}{Carnegie Mellon University, Pittsburgh, Pennsylvania 15213}
\newcommand*{\CMUindex}{5}
\newcommand*{\CUA}{Catholic University of America, Washington, D.C. 20064}
\newcommand*{\CUAindex}{6}
\newcommand*{\SACLAY}{Irfu/SPhN, CEA, Universit\'e Paris-Saclay, 91191 Gif-sur-Yvette, France}
\newcommand*{\SACLAYindex}{7}
\newcommand*{\CNU}{Christopher Newport University, Newport News, Virginia 23606}
\newcommand*{\CNUindex}{8}
\newcommand*{\UCONN}{University of Connecticut, Storrs, Connecticut 06269}
\newcommand*{\UCONNindex}{9}
\newcommand*{\DUKE}{Duke University, Durham, North Carolina 27708-0305}
\newcommand*{\DUKEindex}{10}
\newcommand*{\FU}{Fairfield University, Fairfield CT 06824}
\newcommand*{\FUindex}{11}
\newcommand*{\FERRARAU}{Universita' di Ferrara , 44121 Ferrara, Italy}
\newcommand*{\FERRARAUindex}{12}
\newcommand*{\FIU}{Florida International University, Miami, Florida 33199}
\newcommand*{\FIUindex}{13}
\newcommand*{\FSU}{Florida State University, Tallahassee, Florida 32306}
\newcommand*{\FSUindex}{14}
\newcommand*{\Genova}{Universit$\grave{a}$ di Genova, 16146 Genova, Italy}
\newcommand*{\Genovaindex}{15}
\newcommand*{\GWUI}{The George Washington University, Washington, DC 20052}
\newcommand*{\GWUIindex}{16}
\newcommand*{\Bonn}{Helmholtz-Institut f\"{u}r Strahlen- und Kernphysik, Universit\"{a}t Bonn, Germany}
\newcommand*{\Bonnindex}{17}
\newcommand*{\ISU}{Idaho State University, Pocatello, Idaho 83209}
\newcommand*{\ISUindex}{18}
\newcommand*{\INFNFE}{INFN, Sezione di Ferrara, 44100 Ferrara, Italy}
\newcommand*{\INFNFEindex}{19}
\newcommand*{\INFNFR}{INFN, Laboratori Nazionali di Frascati, 00044 Frascati, Italy}
\newcommand*{\INFNFRindex}{20}
\newcommand*{\INFNGE}{INFN, Sezione di Genova, 16146 Genova, Italy}
\newcommand*{\INFNGEindex}{21}
\newcommand*{\INFNRO}{INFN, Sezione di Roma Tor Vergata, 00133 Rome, Italy}
\newcommand*{\INFNROindex}{22}
\newcommand*{\INFNTUR}{INFN, Sezione di Torino, 10125 Torino, Italy}
\newcommand*{\INFNTURindex}{23}
\newcommand*{\ORSAY}{Institut de Physique Nucl\'eaire, CNRS/IN2P3 and Universit\'e Paris Sud, Orsay, France}
\newcommand*{\ORSAYindex}{24}
\newcommand*{\ITEP}{Institute of Theoretical and Experimental Physics, Moscow, 117259, Russia}
\newcommand*{\ITEPindex}{25}
\newcommand*{\JMU}{James Madison University, Harrisonburg, Virginia 22807}
\newcommand*{\JMUindex}{26}
\newcommand*{\KNU}{Kyungpook National University, Daegu 41566, Republic of Korea}
\newcommand*{\KNUindex}{27}
\newcommand*{\MISS}{Mississippi State University, Mississippi State, MS 39762-5167}
\newcommand*{\MISSindex}{28}
\newcommand*{\UNH}{University of New Hampshire, Durham, New Hampshire 03824-3568}
\newcommand*{\UNHindex}{29}
\newcommand*{\Gatchina}{NRC ``Kurchatov'' Institute, PNPI, Gatchina 188300, Russia}
\newcommand*{\Gatchinaindex}{30}
\newcommand*{\NSU}{Norfolk State University, Norfolk, Virginia 23504}
\newcommand*{\NSUindex}{31}
\newcommand*{\OHIOU}{Ohio University, Athens, Ohio  45701}
\newcommand*{\OHIOUindex}{32}
\newcommand*{\ODU}{Old Dominion University, Norfolk, Virginia 23529}
\newcommand*{\ODUindex}{33}
\newcommand*{\RPI}{Rensselaer Polytechnic Institute, Troy, New York 12180-3590}
\newcommand*{\RPIindex}{34}
\newcommand*{\URICH}{University of Richmond, Richmond, Virginia 23173}
\newcommand*{\URICHindex}{35}
\newcommand*{\ROMAII}{Universita' di Roma Tor Vergata, 00133 Rome Italy}
\newcommand*{\ROMAIIindex}{36}
\newcommand*{\MSU}{Skobeltsyn Institute of Nuclear Physics, Lomonosov Moscow State University, 119234 Moscow, Russia}
\newcommand*{\MSUindex}{37}
\newcommand*{\SCAROLINA}{University of South Carolina, Columbia, South Carolina 29208}
\newcommand*{\SCAROLINAindex}{38}
\newcommand*{\TEMPLE}{Temple University,  Philadelphia, PA 19122 }
\newcommand*{\TEMPLEindex}{39}
\newcommand*{\JLAB}{Thomas Jefferson National Accelerator Facility, Newport News, Virginia 23606}
\newcommand*{\JLABindex}{40}
\newcommand*{\UTFSM}{Universidad T\'{e}cnica Federico Santa Mar\'{i}a, Casilla 110-V Valpara\'{i}so, Chile}
\newcommand*{\UTFSMindex}{41}
\newcommand*{\EDINBURGH}{Edinburgh University, Edinburgh EH9 3JZ, United Kingdom}
\newcommand*{\EDINBURGHindex}{42}
\newcommand*{\GLASGOW}{University of Glasgow, Glasgow G12 8QQ, United Kingdom}
\newcommand*{\GLASGOWindex}{43}
\newcommand*{\VT}{Virginia Tech, Blacksburg, Virginia   24061-0435}
\newcommand*{\VTindex}{44}
\newcommand*{\VIRGINIA}{University of Virginia, Charlottesville, Virginia 22901}
\newcommand*{\VIRGINIAindex}{45}
\newcommand*{\WM}{College of William and Mary, Williamsburg, Virginia 23187-8795}
\newcommand*{\WMindex}{46}
\newcommand*{\YEREVAN}{Yerevan Physics Institute, 375036 Yerevan, Armenia}
\newcommand*{\YEREVANindex}{47}

\newcommand*{\NOWJLAB}{Thomas Jefferson National Accelerator Facility, Newport News, Virginia 23606}
\newcommand*{\NOWHAMPTON}{Hampton University, Hampton, VA 23668}
\newcommand*{\NOWISU}{Idaho State University, Pocatello, Idaho 83209}
\newcommand*{\NOWINFNGE}{INFN, Sezione di Genova, 16146 Genova, Italy}

\author[toASU,toCUA]{P.~Collins}
\author[toASU]{B.G.~Ritchie \corref{cor1}}
\author[toASU]{M.~Dugger}
\author[toCUA,toGWUI]{F.J.~Klein}
\author[toBonn]{A.V.~Anisovich} 
\author[toBonn,JLAB]{E.~Klempt} 
\author[toBonn,toGatchina]{V.A.~Nikonov} 
\author[toBonn,toGatchina]{A.~Sarantsev}

\author[toMISS]{K.P. ~Adhikari}
\author[toFIU]{S. Adhikari}

% ...
\address[toASU]{\ASU}
\address[toBonn]{\Bonn} 
% ...
\address[toCUA]{\CUA} 
% ...
\address[toFIU]{\FIU} 
% ...
\address[toGatchina]{\Gatchina}
% ...
\address[toGWUI]{\GWUI} 
% ...
\address[toMISS]{\MISS} 
% ...

\begin{abstract}
Photon beam asymmetry $\Sigma$ measurements for 
$\omega$ photoproduction in the reaction $\vec{\gamma} p \to \omega p$  are reported 
for photon energies from 1.152 to 1.876  GeV. 
Data were taken using a linearly-polarized tagged photon beam,
a cryogenic hydrogen target, and the CLAS spectrometer in Hall B at Jefferson Lab. 
The measurements obtained markedly
increase the size of the database for this observable, extend
coverage to higher energies, and resolve discrepancies in previously published data. 
Comparisons of these new results with predictions from a chiral-quark-based model and from
a dynamical coupled-channels model indicate the
importance of interferences between $t$-channel meson exchange
and $s$- and $u$-channel contributions, 
underscoring sensitivity to the nucleon resonances included in
those descriptions. 
Comparisons with the Bonn-Gatchina partial-wave analysis
indicate the $\Sigma$ data reported here help to fix the magnitudes of
the interference terms between the leading amplitudes in that 
calculation (Pomeron exchange and the resonant portion of the
$J^P=3/2^+$ partial wave), as well as the resonant
portions of the smaller partial waves with $J^P$= $1/2^-$, $3/2^-$, and $5/2^+$. 
\end{abstract}

\begin{keyword}
meson photoproduction; omega photoproduction;  polarization observable; beam asymmetry 
\PACS 13.60.Le \sep 14.20.Dh \sep 14.20.Gk
\end{keyword}

\cortext[cor1]{Corresponding author}

 \fntext[toNOWJLAB]{Current address: Newport News, Virginia 23606 }
 \fntext[toNOWHAMPTON]{Current address: Hampton, VA 23668 }
 \fntext[toNOWISU]{Current address: Pocatello, Idaho 83209 }
 \fntext[toNOWINFNGE]{Current address: 16146 Genova, Italy }

\date{\today}


\end{frontmatter}

\section{Introduction\label{Intro}}
As a composite system of quarks and gluons, 
the nucleon has an excitation spectrum largely dictated by the underlying
dynamics of the strong interaction. 
Thus, ideally, a description of the excited states of the nucleon 
should arise naturally from a theory built from quantum chromodynamics (QCD). 
However, nearly a half-century of experimental and theoretical study has not produced either a
satisfactory theoretical description or a full empirical inventory of the states present in the nucleon resonance spectrum. 
The current understanding of the shortcomings of these efforts 
(as reviewed in, e.g., 
Refs.~\cite{Tiator:2011pw,Aznauryan:2012ba,Cloet:2013jya,Crede:2013sze, Bazavov:2014xya,Burkert:2016kyi})
could be summarized by
the provocative title of a classic paper from over thirty years ago: 
``Where have all the resonances gone?''~\cite{Koniuk:1979vw}. 
The answer those authors supplied to that question remains part of the current lore today: 
many of the ``missing resonances'' are likely coupled to channels with far smaller strengths
than those states that are coupled to pion-nucleon final states. 

Meson photoproduction has proven to be a 
very productive tool for 
clarifying details of the nucleon resonance spectrum,
complementing other approaches 
in the search for missing resonances. 
Theoretical analyses of the existing database for nucleon excited states 
have usually included Breit-Wigner fits to observables in order to extract masses and widths 
of resonances putatively seen in the experimental results.
Such analyses have shown the nucleon resonance spectrum to
possess many broad and overlapping excitations, 
making progress difficult.
Nonetheless, with respect to the mystery of the missing resonances, 
the current ambiguities in the nucleon resonance spectrum may 
still reflect the fact that the experimental database for the nucleon 
remains dominated by studies of $\pi N$ final states,
as implied in 1980 by Koniuk and Isgur~\cite{Koniuk:1979vw}. 

Over the past two decades, new experimental facilities have become available,
permitting experiments targeting ${\eta} N$, $K \Lambda$, and $K \Sigma$ final states.
These studies have greatly expanded knowlege of nucleon resonances,
as summarized in the reviews noted 
above~\cite{Tiator:2011pw,Aznauryan:2012ba,Cloet:2013jya,Crede:2013sze, Bazavov:2014xya,Burkert:2016kyi}. 
Though many experiments have now investigated photoproduction 
of $\eta N$, $K \Sigma$, and $K \Lambda$ final states, 
the reaction for $\omega$ photoproduction on the nucleon remains relatively unexplored,
even though experiments focused on observables for that reaction 
can address several unique theoretical interests. 
For example, since the threshold for $\omega$ photoproduction (1.108 GeV) 
lies above the thresholds for $\pi$ and $\eta$ photoproduction,
the reaction probes the higher-mass nucleon resonances in 
the so-called third resonance region, where the $\pi N$ and $\eta N$
photoproduction cross sections have become considerably smaller than at lower energies. 
As in $\eta$ photoproduction, the isoscalar nature of the $\omega$  
means that $\omega p$ final states can provide an ``isospin filter'' for the
nucleon resonance spectrum, selecting only isospin $I = \frac{1}{2}$ excitations.
But, in contrast to the spinless isoscalar $\eta$ and $\eta^\prime$ mesons 
and isovector $\pi$ mesons, 
the $\omega$  has an intrinsic spin of 1, yielding a richer set of
angular momentum combinations for intermediate states. 
As a practical matter for experiments, 
the $\omega$ has a much smaller intrinsic width ($\Gamma$=8.49 MeV) 
than the $\rho$ ($\Gamma$=149.1 MeV), 
although both mesons have similar masses~\cite{Olive:2016xmw}. 
The narrower width for the $\omega$ aids greatly in 
identifying that meson in missing-mass reconstructions.
Furthermore, the principal decay mode for the $\omega$ meson
($\pi^+\pi^-\pi^0$ with a branching ratio of 89.2\%~\cite{Olive:2016xmw}) 
includes two charged pions, whose relative ease in detection also facilitates reconstruction.   

All these features of $\omega$ photoproduction have 
stimulated theorists
using a variety of approaches to
harvest information from this particular channel~\cite{
Zhao:1999af,Zhao:2000tb,Zhao:2001qj,Zhao:2002fk,
Oh:2002sq,Oh:2002rb,Shklyar:2004ba,Usov:2006wg,Anisovich:2006bc,
Paris:2008ig,Sarantsev:2008ar,Anisovich:2011fc}. 
Differential cross sections for meson photoproduction form the bulk of the
database for baryon spectroscopy,
and a number of experiments have provided data for 
$\omega$ photoproduction on the 
proton~\cite{BrownHarvardMITPadovaWeizmannInstituteBubbleChamberGroup:1967zz,
ABBHHM:1968aa,Eisenberg:1969xc,Ballam:1972eq,Clifft:1977yi,Friman:1995qm,Klein:1998xy,
Barth:2003kv,Battaglieri:2002pr,Ajaka:2006bn,Williams:2009ab,Sumihama:2009gf,Wilson:2015uoa}.  
However, differential cross sections alone are insufficient to deconvolute the nucleon resonance spectrum. 
Polarization observables in meson photoproduction, 
where selection of the orientations of the initial spins of the nucleon and photon,
as well as measurements of the orientation of the intrinsic angular momentum of particles in the final state, 
give additional insight into the details of the 
reaction mechanism~\cite{Barker:1975bp,Kloet:1998js,Roberts:2004mn}. 
Such observables can arise from interferences between contributing amplitudes, 
consequently demanding much more specificity about the properties of the
hypothesized resonance states 
involved in the reaction than the differential cross sections. 

The photon beam asymmetry $\Sigma$ is one such polarization observable.
As discussed in Ref.~\cite{Pichowsky:1994gh},
$\Sigma$ for vector meson photoproduction on a nucleon
is obtained with a linearly-polarized photon beam incident on an unpolarized target. 
Using a coordinate system where the $z$-axis is defined by the 
incoming photon direction, this observable 
can be expressed in the center-of-mass frame as 
\begin{equation} \label{eq:Sigmadef}
	\frac{d\sigma}{d\Omega} = 
		\frac{d\sigma_0}{d\Omega} \left[1 - P_\gamma \Sigma \cos \{ 2 \left ( \varphi-\alpha \right) \} \right],
\end{equation}
where $\frac{d\sigma}{d\Omega}$ is the differential cross section for the reaction
using a polarized photon beam, 
 $\frac{d\sigma_0}{d\Omega}$ is the {\em{unpolarized}} differential cross section, 
$P_\gamma$ is the degree of linear polarization of the photon beam, 
$\varphi$ is the azimuthal angle of the photoproduced meson
relative to a plane parallel to the floor in the laboratory frame,
and $\alpha$ is the azimuthal angle between the photon beam polarization plane
and the laboratory floor plane. 
Predictions of $\Sigma$ show that this observable is very sensitive to the
details of which resonances are involved in the 
$\vec{\gamma} p \to \omega p$ 
reaction~\cite{Zhao:1999af,Zhao:2000tb,Paris:2008ig,Shklyar:2004ba}.
However, the three published sets of measurements of $\Sigma$ for this 
reaction~\cite{Ajaka:2006bn,Klein:2008aa,Vegna:2013ccx}, 
which have yielded a total of 74 data points,
are often in conflict with each other.

The results of the experiment described in this report markedly
increase the database for $\Sigma$
by adding nearly four times more data points and extending
coverage to higher incident photon energies. 
These new data possess finer energy and angle resolution, 
and resolve discrepancies among the previously published results. 
The measurements reported here also should motivate new theoretical analyses of this reaction 
that will further clarify the nucleon resonance spectrum.
	
\section{Experiment\label{experiment}}
This experiment was
part of a program carried out in Experimental Hall B 
at the Thomas Jefferson National Accelerator Facility (Jefferson Lab)
in order to provide the  large set of observables for exclusive meson photoproduction 
needed to better understand the nucleon resonance spectrum. 
The results reported here are based on analyses of data taken 
during the ``g8b'' running period
at that facility. 
Data for $\Sigma$ for $\pi^+$, $\pi^0$, $\eta$, and $\eta^\prime$ photoproduction
from the same running period
were extracted and reported previously~\cite{Dugger:2013crn,Collins2017,CLASg8b}.
Those  publications provide details of the experiment and running conditions, 
which we summarize here. 

The linearly-polarized photon beam was generated by coherent bremsstrahlung
of the primary Jefferson Lab electron beam incident on a diamond radiator~\cite{Bilokon:1982wu}.
In coherent bremsstrahlung, the normal bremsstrahlung spectrum
is enhanced at specific photon energies due to the lattice excitations 
within the oriented crystal.  
Intensity enhancements in the resulting photon spectrum above
the normal bremsstrahlung spectrum possess significant polarization. 
The peak corresponding to the highest-energy polarized photons is called the coherent edge. 
By adjusting the orientation of the diamond radiator with respect to the incident electron beam, 
the polarization vector of the photon beam
can be rotated and the energy of the coherent edge can be adjusted.

For this work, a 4.55-GeV electron beam scattering from the crystal planes 
of a 50-$\mu m$-thick diamond radiator
produced the linearly-polarized photons,
with a remotely-controlled goniometer used to adjust the polarization direction.
The degree of linear polarization for the photon beam was estimated with a calculation 
that used knowledge of the goniometer orientation and the geometry of the full beam line, 
including the degree of collimation of the photon beam,
during each portion of the data collection period
\cite{Livingston:2011}. 
Timing information and energy definition for the polarized photons were
obtained using the bremsstrahlung photon tagger in Hall B~\cite{Sober:2000we}.

The photon beam impinged on a 40-cm-long liquid hydrogen target
located 20 cm upstream of the center of the CEBAF Large Acceptance
Spectrometer (CLAS)~\cite{Mecking:2003zu}.
CLAS consists of six (ideally) identical charged-particle magnetic spectrometers
contained within a superconducting toroidal magnet 
that generates an approximately azimuthal magnetic field distribution. 
Information from the CLAS subsystems 
was analyzed to provide four-momentum and reaction vertex information  
event-by-event for each charged particle originating in the cryogenic target region 
and passing through the tracking regions of the spectrometer.  
For these measurements,
data from the drift-chamber subsystem 
for tracking charged particles~\cite{Mestayer:2000we}, the time-of-flight subsystem~\cite{Smith:1999ii}, 
and a plastic scintillator array surrounding the cryogenic target 
(which determined when charged particles passed from the target into the 
drift chamber region)~\cite{Sharabian:2005kq},
provided the primary  information for determining the four-momenta
of the recoil proton, the photoproduced mesons, and their decay products. 

The running period was split into intervals of approximately one week
during which data were taken at a specific setting of the coherent-edge energy.
The coherent-edge settings were 1.3, 1.5, 1.7, and 1.9 GeV. 
Approximately every hour within each interval, the polarization plane for the $\vec{E}$-field 
of the photon beam was adjusted to either be parallel to the floor 
or perpendicular to the floor in order to minimize the effects of 
any change in the CLAS acceptance within a coherent-edge setting.  
Measurements where the same photon energy was present in adjacent coherent-edge settings 
were used to check for consistency in polarization estimates for a given photon energy.
These consistency checks indicated that the uncertainty in the photon beam polarization was 6\%,
as noted in Ref.~\cite{Dugger:2013crn}.
Runs with an amorphous carbon radiator were also taken periodically during each interval
to provide data for unpolarized photons. 
The use of the amorphous radiator and alternate polarization orientations 
reduced systematic uncertainties
arising from the non-uniform CLAS acceptance for the charged decay products of the $\omega$ meson
through  the comparison of combinations of the data from the different polarization
orientations with data from the amorphous radiator. 

\section{Data reduction and analysis}
For each charged-particle track seen in CLAS,
the measured speed $v$, $\beta=v/c$, and three-momentum 
were used for particle identification 
via the GPID algorithm~\cite{GPID, Dugger:2013crn,Collins2017}.
That algorithm compares the measured $\beta$ of the particle whose identity is to be determined with 
estimated values of $\beta$ based on hypothetical identities for that particle. 
The hypothetical particle identity that provided a $\beta$ value closest to the 
measured $\beta$ was then assigned to that particle. 
A visualization of the performance of this technique may be seen in
Fig. 1 of Ref.~\cite{Dugger:2013crn}.

With the identity of each scattered particle established,
corrections were made for the energy lost by each scattered charged particle 
as that particle passed through the
materials in the cryogenic target and the CLAS detector with the CLAS ELOSS program \cite{ELOSS}.  
The timing information for each charged-particle track
was used to determine the time when that track originated in the target (i.e., the vertex time).
Independently, the timing information of each electron 
detected at the focal plane of the tagger 
was used to determine the time when the corresponding bremsstrahlung photon arrived at the vertex
(photon time). 
The photon whose photon time most closely matched with the vertex time was selected as
the photon that caused the reaction. 
This selection was important because 
the intensity of the electron beam impacting the radiator of the photon tagger 
was such that multiple photons 
could arise from a single pulse of beam electrons. 
Events were rejected where an additional photon was within $\pm$ 1 ns of the selected photon 
in order to avoid any ambiguity in determining which photon caused the reaction.

The kinematic quantities determined from the time-of-flight and drift-chamber systems 
yielded good definition of the four-momenta 
for the $p$, $\pi^+$, and $\pi^-$ particles scattered into CLAS. 
Nonetheless, to simultaneously correct for imperfections 
in the map for the magnetic field of CLAS, 
momentum corrections for tracks were determined 
by demanding four-momentum conservation
 in a kinematic fit of a large sample of $\gamma p \rightarrow \pi^+\pi^- p$ events 
seen in the spectrometer where all three final-state particles were detected, 
in the same manner as discussed in Ref.~\cite{CLASg8b}.

\begin{figure}[ht]
\centerline{\includegraphics[width=\columnwidth]{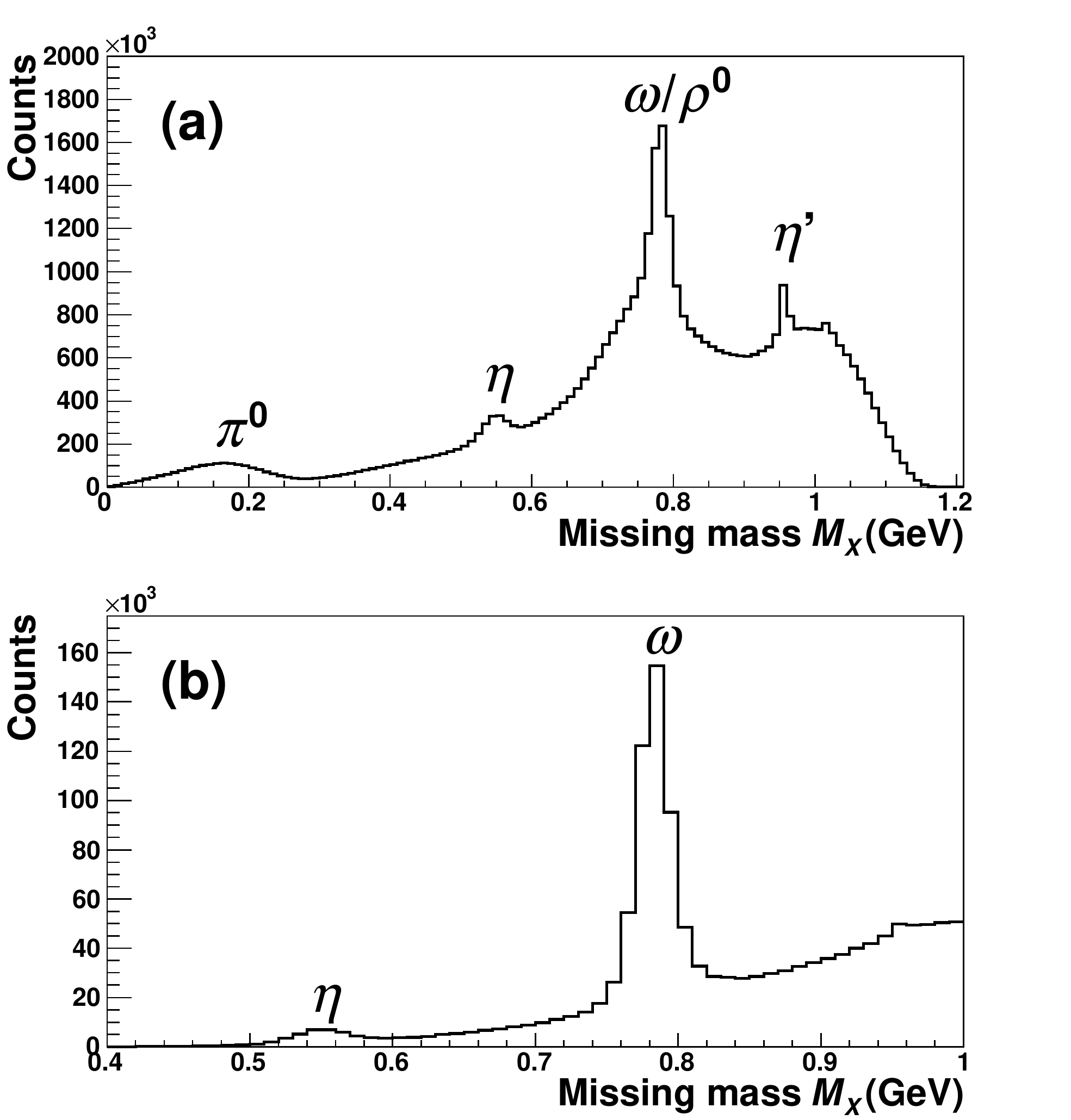}}
\caption{Missing mass $M_X$ for the reaction $\gamma p \to pX$  
for the 1.9-GeV coherent-edge setting,
with peaks corresponding to various photoproduced mesons.
Upper panel: Full missing-mass spectrum.
Lower panel: Missing-mass spectrum for the reaction $\gamma p \to pX$
requiring $X \rightarrow \pi^-\pi^+\pi^0$, as described in the text.
\label{massPlots}}
\end{figure}

Based on the assumption that the reaction observed was $\gamma p \to pX$,
the polar scattering angle and the magnitude of the three-momentum 
for the proton recoiling from meson photoproduction 
can be used to calculate the mass $M_X$ of the missing state $X$. 
As seen in the upper panel of Fig. \ref{massPlots}, however,
a sizeable background in the missing-mass spectrum for $\gamma p \to pX$
appears under the 
peak associated with photoproduction of the $\omega$ meson
due to multi-pion 
and $\rho$ meson photoproduction. 
This background was reduced by requiring
that the recoil proton and charged pions resulting from the decay 
$\omega \rightarrow \pi^+\pi^-\pi^0$ were detected in CLAS, 
and then identifying a neutral meson by assuming 
the decay $\omega \rightarrow \pi^+\pi^-Y$ with 
the restriction on the missing mass $M_Y=M(\pi^0)$.
This requirement effectively removed contributions from $\rho$ photoproduction
and significantly reduced the background beneath the photoproduced $\omega$ peak, 
as exemplified in the lower panel of Fig. \ref{massPlots}. 
The remaining background is attributable primarily to multi-pion photoproduction,
which was removed to extract the $\omega$ yield as described below. 

\begin{figure}[ht]
\centerline{\includegraphics[width=\columnwidth]{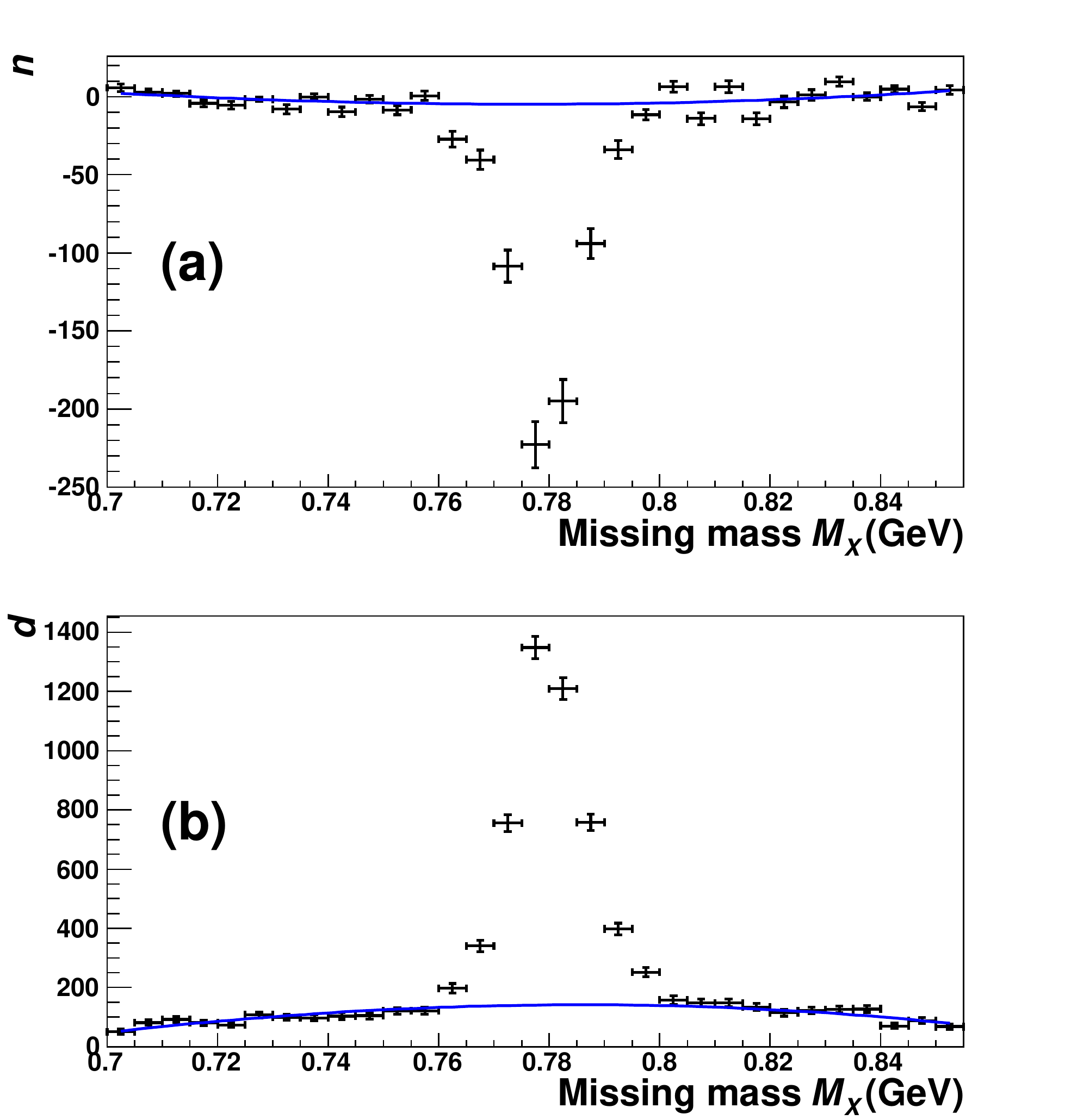}}
\caption{Example of yield extraction for the numerator $n$ (upper panel) 
and denominator $d$  (lower panel)
defined by Eq.{~\protect\ref{eq:Sigma_moments}}
for $E_\gamma$=1.314 GeV 
and  $\cos(\theta^\omega_{c.m.})=-0.55$.
The solid line indicates the background fit discussed in the text.
\label{MomPlot}}
\end{figure}

These $\omega$ missing-mass spectra were then 
split into 27-MeV-wide bins in photon energy.
Incorporating the results of the kinematic fit described above, 
the centroids for each of the photon energy bins used in this analysis
were determined to an accuracy that was typically better than $\pm$ 0.1 MeV, 
and always better than $\pm$ 0.5 MeV. 
These binned spectra were then analyzed
as  in Refs.~\cite{Dugger:2013crn,Collins2017} with
a Fourier-moment method to extract the 
beam asymmetry as a function of the center-of-mass 
meson scattering angle $\cos(\theta^\omega_{c.m.})$ for
the specific incident photon energy $E_\gamma$ bins (and, consequently,
center-of-mass $W$ bins) chosen for this analysis.
Cosine-$n\varphi$-moment histograms (where $n$=0, 2, 4) were constructed 
by taking each $\omega$ event in the missing-mass histograms 
and weighting that event by the value of $\cos n\varphi$ corresponding to that event. 
With this approach, events within a particular $\cos(\theta^\omega_{c.m.})$ bin 
for $\varphi$ are combined simultaneously to determine $\Sigma$. 
Applying this Fourier-moment method to $\Sigma$, 
the resulting equation for the beam asymmetry may be written as
\begin{equation}\label{eq:Sigma_moments}
\Sigma=\frac{\tilde{Y}_{\perp 2} - \tilde{Y}_{\para 2}}
{\frac{P_\para}{2}(\tilde{Y}_{\perp 0} + \tilde{Y}_{\perp 4}) + \frac{P_\perp}{2}(\tilde{Y}_{\para 0} + \tilde{Y}_{\para 4})} \ , 
\end{equation}
where  $\tilde{Y}_{\perp n}$($\tilde{Y}_{\para n}$) is the background-subtracted meson yield 
for a photon beam with polarization vector perpendicular (parallel) 
to the laboratory floor, 
normalized by the number of incident photons for that particular polarization orientation,
with each event weighted 
according to the Fourier moment $\cos  n \varphi $, 
and $P_\perp$( $P_\para$) is the degree of photon polarization. 
The numerator and denominator in Eq.~\ref{eq:Sigma_moments}
were constructed for each kinematic bin. 
For each Fourier-moment histogram within a kinematic bin,
the meson yield was determined by removing the background under the $\omega$
meson peak.
Since the $\omega$ peak shows up very clearly in such histograms, 
this background subtraction was accomplished by fitting each of the Fourier-moment histograms
with the combination of a second-order polynomial shape 
(describing the multi-pion background)
and a Gaussian (describing the $\omega$ peak). 
The background was well-described by the polynomial shape,
and the uncertainty in the background was given by
the uncertainty in this shape determined by the fit. 
The $\omega$ yield for that particular histogram 
was then obtained by subtracting the polynomial from the
original distribution and then summing the remaining events
within the area of the $\omega$ peak. 
Examples of Fourier-moment histograms with background fits are shown in Fig.~\ref{MomPlot}.
Finite-size bin corrections for extracted $\Sigma$ values
are also addressed with this method, 
as described in Ref.~\cite{Dugger:2013crn}. 
Using preliminary estimates of the spin-density matrix elements for $\omega$ decay \cite{Vernarsky:2014vaa}, 
the values of $\Sigma$ also were corrected to account for CLAS acceptance variations 
for charged pions due to any polarization transferred to those pions in the $\omega$ decay. 


\section{Statistical and systematic uncertainties \label{sec:Unc}}

A statistical uncertainty for $\Sigma$ was determined for each kinematic bin, 
based on the definition in Eq.~\ref{eq:Sigma_moments}.
With that definition, many experimental quantities 
(e.g., acceptance, target thickness, single-particle detection 
efficiencies) canceled in Eq.~\ref{eq:Sigma_moments} 
so that the uncertainty for a particular kinematic bin 
was to a great extent dictated by statistical uncertainties 
in the $\omega$ yield for that bin~\cite{Dugger:2013crn,Collins2017}.
The relative normalization of the photon flux for 
the different coherent-edge settings and 
polarization orientations had statistical uncertainties much less than 1\%,
contributing negligibly to the overall uncertainty in $\Sigma$.
The effects arising from polarization transfer to 
the charged pions from $\omega$ decay used to reconstruct the $\omega$
provided an additional statistical uncertainty in $\Sigma$ for each kinematic bin,
which was conservatively set to 0.01 for all kinematic bins based on
those simulations.
This uncertainty due to polarization transfer was then added in quadrature to 
the statistical uncertainty in $\Sigma$ from Eq.~\ref{eq:Sigma_moments}
to arrive at the overall statistical uncertainty for that kinematic bin. 

The systematic uncertainties for all values of $\Sigma$ obtained
at a particular photon energy $E_\gamma$  
arose from the uncertainties in the polarization of the photon beam 
and the relative flux normalization for that particular photon energy.
By analyzing $\Sigma$ measurements for the same $E_\gamma$ taken at different coherent-edge settings, as noted above, 
the systematic uncertainty in the photon beam polarization 
for a particular polarization orientation was found to be 4\%,
as reported in Ref.~\cite{Dugger:2013crn}.
Since two different polarization orientations are combined 
to obtain $\Sigma$, and the photon-beam flux contributions are negligible,
adding the two polarization uncertainties in quadrature resulted in an 
estimated systematic uncertainty in $\Sigma$ of 6\% for all photon energies,
as given in Refs.~\cite{Dugger:2013crn,Collins2017}.

\begin{figure*}
\centerline{\includegraphics[width=160mm]{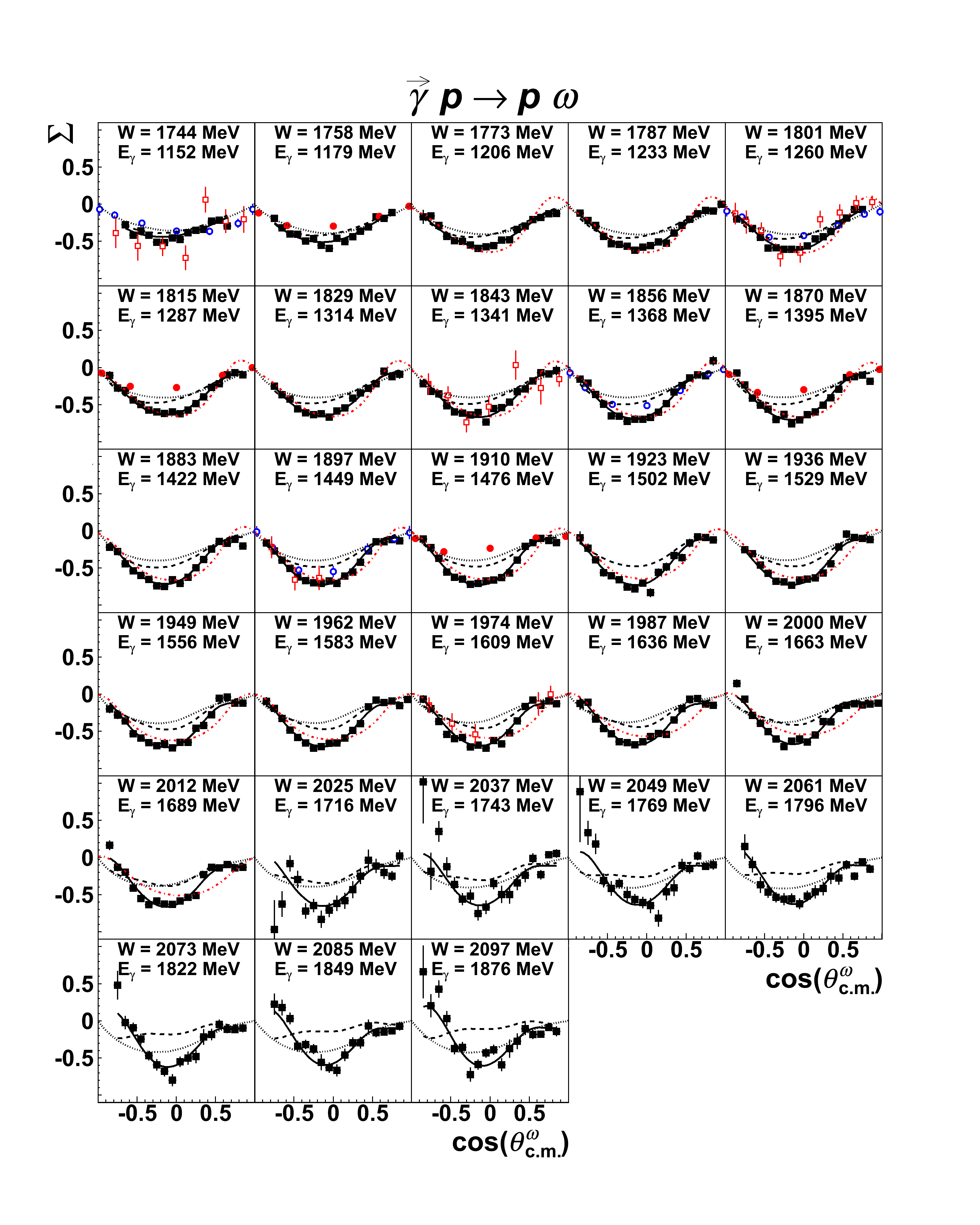}}
\caption{ (Color online) Photon beam asymmetry $\Sigma$ as a function of $\cos(\theta^{\omega}_{c.m.})$ 
for $\vec{\gamma} p \to \omega p$ for $E_\gamma$   from 1.152 GeV  
($W$=1.744 GeV) to 1.876 GeV  ($W$=2.097 GeV). 
Uncertainties for data reported here (black squares) are statistical for each point shown,
and do not include a 6\% systematic uncertainty at each energy. 
Also shown are results from CBELSA/TAPS (open red squares~\protect{\cite{Klein:2008aa}})
and GRAAL (blue open circles~\protect{\cite{Ajaka:2006bn}}
and red solid circles~\protect{\cite{Vegna:2013ccx}}).
The predictions described in the text by CQM (black dotted lines) and CCA (red dotted-dashed lines), 
as well as new fits with the Bonn-Gatchina formalism 
discussed in the text (black dashed and black solid lines), are also shown.}  
\label{sigmadists}
\end{figure*}

\section{Results and comparisons with prior results}

The results for $\Sigma$ obtained here \cite{CLASdatabase} 
are shown in Fig.~\ref{sigmadists}.
The uncertainty shown at each point is 
the statistical uncertainty in $\Sigma$ described in Sect.~\ref{sec:Unc}.
Additionally, the entire $\Sigma$ distribution at each center-of-mass energy $W$ 
shown in Fig.~\ref{sigmadists} possesses
the 6\% systematic uncertainty 
due to the uncertainty in the polarization of the photon beam noted in Sect.~\ref{sec:Unc}.

The published results  for $\Sigma$ from
the CBELSA/TAPS collaboration~\cite{Klein:2008aa} and the
GRAAL collaboration~\cite{Ajaka:2006bn,Vegna:2013ccx}
are also shown in Fig.~\ref{sigmadists}.
The data reported here extend knowledge
of this observable well beyond the $W$ range studied in those previous experiments,
and generally have higher precision.
The energy-bin width here (typically $\sim$27 MeV)
is considerably smaller than that for the previous studies
(typically $\sim$100 MeV).
Near threshold where observables change rapidly,
the better energy resolution of the current study will
prove very useful to future analyses of the nucleon resonance spectrum.
Angular resolution is also much better in the current
results, which will enable more detailed comparisons
with the theoretical predictions below. 
   
As is clear from Fig.~\ref{sigmadists}, our results and
the previous measurements of $\Sigma$ 
indicate that this observable is negative and
significantly different from zero near $\cos(\theta^\omega_{c.m.})$=0.
This general observation will be of interest to the discussion 
in the next section. 
With respect to the shape of $\Sigma$ as a function of $\cos(\theta^\omega_{c.m.})$,
all measurements suggest
a generally similar angular dependence for $\Sigma$.
In more detail, 
the overall agreement with the CBELSA/TAPS results~\cite{Klein:2008aa} is good
for all energies reported in that work, 
though the experimental uncertainties are much larger for
that earlier work than the uncertainties in 
the results reported here. 
On the other hand, the results reported here
generally do not agree with those from the GRAAL publications
except at the most forward and most backward angles
where $\Sigma$ approaches zero. 
The older results from GRAAL~\cite{Ajaka:2006bn} generally
are smaller in magnitude than the data reported here.  
The data reported here also disagree
with the newer GRAAL measurements~\cite{Vegna:2013ccx}
at intermediate angles by a factor of about 2. 
Furthermore, at about $E_\gamma\ge$1.3 GeV, the two GRAAL publications
appear to disagree with each other at intermediate angles.
The results from both GRAAL publications generally disagree with the results from CBELSA/TAPS
at most intermediate angles.
Since the angular dependence of $\Sigma$ 
in both GRAAL datasets appears similar to that measured here
and by CBELSA/TAPS, 
and since the CBELSA/TAPS results agree with the results reported here,
the observed discrepancies between our results and GRAAL  
may be due to an unknown systematic effect in the yield extraction
for the more recent GRAAL publication. 
Regardless, our data weigh in favor of the CBELSA/TAPS results versus the GRAAL measurements.

\section{Comparison with theoretical predictions}
When coupled with measurements
of other observables and/or other reactions, 
these new data can provide important tests and constraints for theoretical predictions
of the nucleon resonance spectrum.
As examples for this discussion, three different approaches
are compared with these new $\Sigma$ data.
In these comparisons, the reader should note that
large $\Sigma$ values at intermediate angles 
are not produced by $t$-channel exchange,
but rather must arise from contributions 
from the $s$- and $u$- channels~\cite{Zhao:1999af,Zhao:2000tb,Paris:2008ig}.
This makes measurements at those angles especially useful in testing
contributions of nucleon resonances. 

The first set of predictions provided here (denoted CQM hereafter) 
used a $SU(6) \otimes SU(3)$ constituent-quark model  
with an effective chiral  Lagrangian
approach for the reaction dynamics~\cite{ZhaoPC,Zhao:1999af,Zhao:2000tb,Zhao:2001qj,Zhao:2002fk,Li:1997gd}. 
The Lagrangian for the quark-$\omega$
coupling utilized a non-relativistic constituent-quark 
treatment for the nucleon.
Thus,
as suggested in Sect.~\ref{Intro}, such an approach is
an approximation to the idealized fundamental quark-level description
of the nucleon in terms of QCD. 
The CQM consisted of three pieces: 
$s$- and $u$-channel nucleon exchange, 
$t$-channel Pomeron (natural parity) exchange,
and $t$-channel $\pi^0$ (unnatural parity) exchange. 
In this model, with respect to the comment above
concerning large values of $\Sigma$ at intermediate angles,
those large asymmetries cannot be produced by Pomeron and/or $\pi^0$ exchanges,
but rather must arise through the interference between (1) the Pomeron and/or $\pi^0$ exchange
and (2) the ($s$- and $u$-channel) effective Lagrangian nucleon exchange. 
The tree-level diagrams were calculated explicitly in this approach, 
and the quark-model wavefunctions for the nucleons and baryon resonances
provided a form factor for the interaction vertices. 
Consequently, all of the $s$- and
$u$-channel resonances could be consistently included 
to facilitate searching for ``missing resonances." 
A set of eight well-known resonances 
expressed in terms of their representations in $SU(6) \otimes SU(3)$
were included. 
Specifically, those resonances were the 
$N(1440)1/2^+$, $N(1520)3/2^-$, $N(1535)1/2^-$, $N(1680)5/2^+$, 
$N(1710)1/2^+$, $N(1720)3/2^+$, $N(1900)3/2^+$,  and $N(2000)5/2^+$. 

The CQM predictions~\cite{ZhaoPC} are compared with the results from this and prior experiments
in Fig.~\ref{sigmadists}. 
As noted above, 
any large observed asymmetry in the CQM results must arise through interference effects
between Pomeron and/or $\pi^0$ $t$-channel exchange and the $s$- and $u$-channel 
contributions.  
Hence, the general observation above 
that all measurements of $\Sigma$ in Fig.~\ref{sigmadists}
are significantly different from zero at intermediate angles
implies that such  interferences indeed are present and are critical to understanding the
data for all energies studied here.
Next, we note that the parameters in CQM were adjusted~\cite{ZhaoPC} to fit 
the older GRAAL data~\cite{Ajaka:2006bn}.
The CQM predictions therefore do not consider more recent data.
Thus, as expected from the discussion of previously published data,
where the older GRAAL data exist, 
the CQM predictions markedly underpredict the data reported here.
However, simply multiplying the predictions by a factor of 2 at those energies
results in a prediction very close in magnitude
and shape to the data reported here.
This suggests that the resonances used and the interferences found in
Refs.~\cite{Zhao:1999af,Zhao:2000tb,Zhao:2001qj,Zhao:2002fk}
could still be mostly correct, 
aside from the constants used to fix the magnitudes of
the various terms to fit the older GRAAL data.
The CQM calculation indicated that the 
$N(1720)3/2^+$ played 
the major role in the shape of the angular dependence seen for $\Sigma$ near
threshold to about $W$=1.9 GeV. 
Given the interest in quark-based descriptions of the nucleon
and the suggestive agreement seen 
between the shape of the angular dependence in the CQM predictions and
this new data, an updated fit with this approach would be interesting. 

A second set of predictions for $\Sigma$
used a coupled-channels 
approach (denoted CCA hereafter) where pion- and photon-induced 
reactions were considered simultaneously
while unitarity was preserved~\cite{Paris:2008ig}.
We were provided predictions for $E_\gamma$=1.206 to 2.012 GeV~\cite{ParisPC}.
The CCA formalism was one component of a program of analyses for
electromagnetic meson production data at Jefferson Lab's 
Excited Baryon Analysis Center. 
The CCA predictions for $\pi$ and $\omega$ production
employed a set of six intermediate channels
($\pi N$, $\eta N$, $\pi \Delta$, $\sigma N$, $\rho N$, $\omega N$), and
incorporated off-shell effects using the dynamical coupled-channel method 
developed by Matsuyama, Sato, and Lee~\cite{Matsuyama:2006rp}.
A large value for $\Sigma$ 
in this model required $s$- and $u$-channel
contributions in one or more partial-waves for the $\omega N$ intermediate state. 
The incorporation of the $\omega N$ intermediate state 
was found to produce marked changes
in the $D_{15}$ partial-wave amplitude,
underscoring the importance of considering 
multiple channels and reactions simultaneously. 
The predictions for the unpolarized differential cross sections
for $\gamma p \to \omega p$ that were incorporated in the fit
were in very good agreement with published data~\cite{Barth:2003kv}.

Though a large database of about 1800 data points was incorporated in the CCA fit,
no $\Sigma$ data for $\vec{\gamma} p \to \omega p$ were included in the fit.
The CCA results thus represented a true prediction for that observable.
When the CCA predictions of Ref.~\cite{Paris:2008ig} were compared to 
the only $\Sigma$ data existing at that time (the older GRAAL
data~\cite{Ajaka:2006bn}),
the calculation predicted $\Sigma$ with a magnitude greater than those data
except at the lowest energy. 
By contrast, as seen in Fig.~\ref{sigmadists}, 
the same CCA predictions agree well with the new data reported here~\cite{Paris:2008ig,ParisPC}. 
Given that CCA simultaneously fits six different
channels with good success, 
the agreement seen in Fig.~\ref{sigmadists} suggests the model likely is correct
in much of its description of the underlying dynamics
for the various reaction channels.
While the CQM model indicated that the $N(1720)3/2^+$ resonance
was critical to understanding $\Sigma$ near threshold,
CCA found instead that the $D_{13}$ partial-wave amplitude,
using resonances with bare masses of 1.899 GeV and 1.988 GeV,
were the most significant component in generating the asymmetry. 
Given the importance of multi-channel coupling effects,
the disagreement between CQM and CCA perhaps lies in a failure of the CQM
calculations to respect unitarity in the $\gamma p$ reactions.
Updating the CCA work with this new data could clarify 
resonance contributions for $W<$ 2 GeV. 

A third set of calculations was developed 
for this publication by the Bonn-Gatchina group
(denoted here as BG) 
from their partial-wave analysis~\cite{Anisovich:2006bc,Anisovich:2011fc, SarantsevPC}.
The BG analysis incorporates a large database (more than 2000 data points) 
of differential cross sections and spin observables from
pion- and photon-induced reactions on the nucleon. 
This approach makes use of dispersion relations 
based on the N/D technique, 
corresponding to the solution of the Bethe–Salpeter equation with separable interactions.
Consistent in part with the CQM results noted above,
a recent analysis with the BG approach of $\omega$ photoproduction
using newly-available data on 
the $E$ and $G$ spin observables 
found that the $J^P$=$3/2^+$ partial wave provided
the strongest contributions to describing correctly 
the $W$ behavior of spin observables~\cite{Denisenko:2016ugz}.
The Pomeron exchange contribution amounted to nearly
half the total cross section for the $\gamma p \rightarrow \omega p$
at $W$=2 GeV.

Fig.~\ref{sigmadists} shows the results from these new fits 
using the BG approach with (black solid line)
and without (black dashed line) considering the data reported here,
indicating the impact of these new data on the BG parameters.
As illustrated by the differences seen in that figure between the original BG predictions
and those considering the data reported here,
our new $\Sigma$ data significantly change the BG predictions
at all energies, with the differences becoming more pronounced as $W$ increases.
When incorporated into the new fits, 
the $\Sigma$ data reported here helped refine details of
the interference between the leading amplitudes in the 
calculation -- the Pomeron exchange and the resonant portion of the
$J^P=3/2^+$ partial wave -- as well as the resonant
portions of the smaller partial waves 
(e.g., the $J^P$= $1/2^-$, $3/2^-$, and $5/2^+$).
Further details and discussion of this new BG analysis will form the subject 
of a subsequent publication. 

\section{Conclusions}
The results of the experiment reported here have
provided hundreds of new high-precision data 
for $\Sigma$  for the reaction $\vec{\gamma} p \to \omega p$,
nearly quadrupling the size of the database for this observable.
The results resolve the disagreements between
the previously published datasets from CBELSA/TAPS and GRAAL,
where agreement between the CBELSA/TAPS results and the results reported here
would suggest that the GRAAL measurements 
are systematically too small. 
Our results also extend the database for $\Sigma$  to higher energies, 
facilitating explorations of nucleon excitations within that energy regime.
An initial study with the BG partial-wave approach
indicates these new data significantly impact the predictions for $\Sigma$
using that approach, showing that these new data will help further refine 
understanding of the interference of many of the partial-wave amplitudes that
contribute to the the $\omega p$ photoproduction reaction. 
Similar comparisons with other theoretical models 
will further enhance our knowledge of the nucleon resonance spectrum. 

\section{Acknowledgements}
The authors express appreciation to Q. Zhao and M. Paris
for providing predictions for $\Sigma$ shown in Fig.~\ref{sigmadists}.
The authors gratefully acknowledge the work of the Jefferson Lab staff
and financial support by the U.S. National Science Foundation,
the Deutsche Forschungsgemeinschaft, 
the Russian Science Foundation,
the Chilean Comisi\'on Nacional de Investigaci\'on Cient\'ifica y Tecnol\'ogica (CONICYT),
the Italian Istituto Nazionale di Fisica Nucleare,
the French Centre National de la Recherche Scientifique,
the French Commissariat \`{a} l'Energie Atomique,
the Scottish Universities Physics Alliance (SUPA),
the United Kingdom's Science and Technology Facilities Council,
and the National Research Foundation of Korea. 
This material is based  upon work supported by the U.S. Department of Energy, Office of Science,
Office of Nuclear Physics under contract DE-AC05-06OR23177.
The Southeastern Universities Research Association (SURA) operates the
Thomas Jefferson National Accelerator Facility for the United States
Department of Energy under contract DE-AC05-84ER40150.

\section{References}

\bibliography{ref-tilde}{}

\end{document}